\documentclass[aps,reprint,showpacs,superscriptaddress,groupedaddress]{revtex4-1}  % for double-spaced preprint
\usepackage{graphicx}  % needed for figures
\usepackage{dcolumn}   % needed for some tables
\usepackage{bm}        % for math
\usepackage{amssymb}   % for math

\newcommand\figref[1]{Figure~\ref{#1}}
\newcommand{\subfigref}[2]{Figure~\ref{#1}(#2)}
\newcommand{\eqref}[1]{(\ref{#1})}

\begin{document}

\title{Tunable chiral spin texture in magnetic domain-walls}

\author{J.~H.~Franken}
\author{M.~Herps}
\author{H.~J.~M.~Swagten}\email[]{h.j.m.swagten@tue.nl}
\author{B.~Koopmans}
\affiliation{Department of Applied Physics, Center for NanoMaterials, Eindhoven University of Technology, P.O.~Box~513,
5600 MB Eindhoven, The Netherlands}

\maketitle

\textbf{Magnetic domain-walls (DWs) with a preferred chirality exhibit very efficient current-driven motion \cite{Thiaville2012,Emori2013a,Ryu2013}. Since structural inversion asymmetry (SIA) is required for their stability, the observation \cite{Je2013} of chiral domain walls in highly symmetric Pt/Co/Pt is intriguing. Here, we tune the layer asymmetry in this system and observe, by current-assisted DW depinning experiments, a small chiral field which sensitively changes. Moreover, we convincingly link the observed efficiency of DW motion to the DW texture, using DW resistance as a direct probe for the internal orientation of the DW under the influence of in-plane fields. The very delicate effect of capping layer thickness on the chiral field allows for its accurate control, which is important in designing novel materials for optimal spin-orbit-torque-driven DW motion.}

\begin{figure*}[tbhp]
\includegraphics[width= \linewidth]{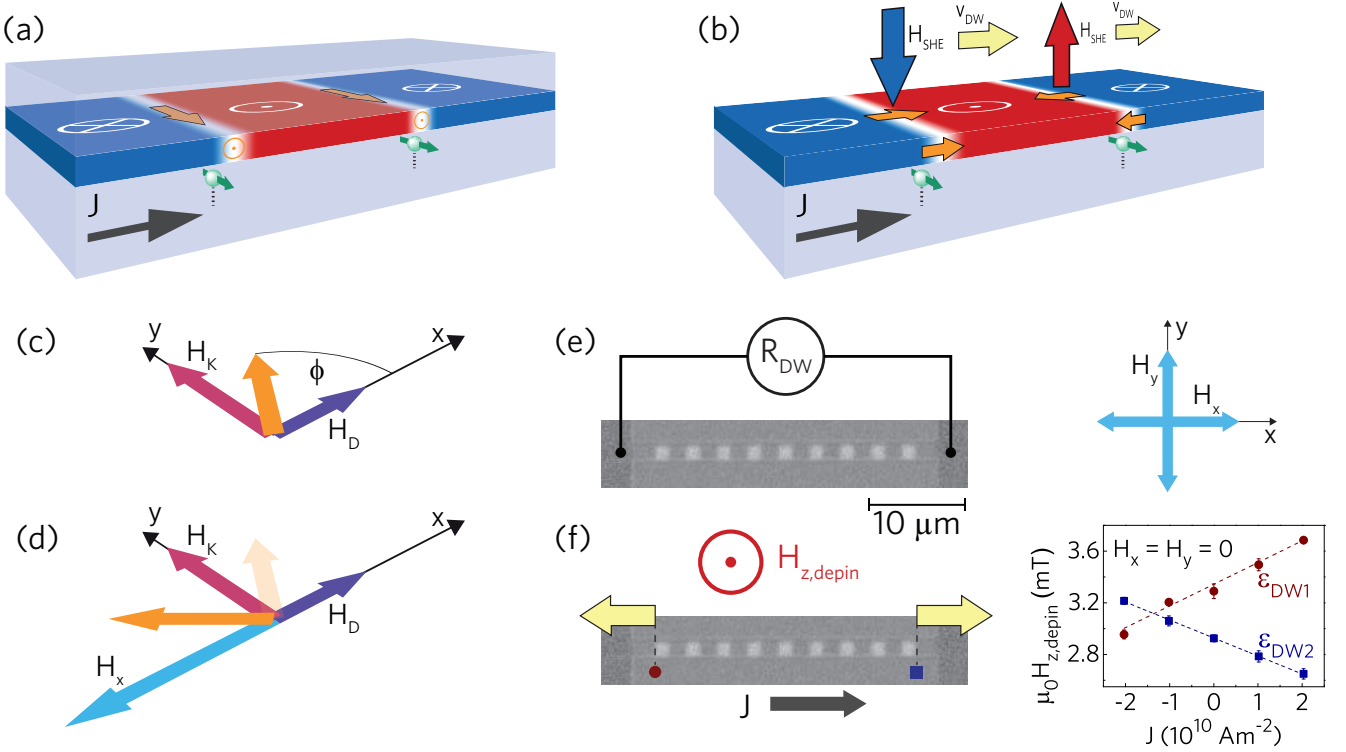} \caption{\label{fig:dmi:Figure1} Tuning the chirality of magnetic DWs. (a) In a symmetric layer system, Bloch walls are favored (orange arrows), on which the spin Hall current (green spins) cannot exert a torque. (b) If SIA is introduced, chiral N\'eel walls with alternating orientation become stable, which are moved uniformly (yellow arrows) by the effective spin Hall field $H_{\rm{SHE}} \propto \cos \phi$. (c) The in-plane DW angle $\phi$ (orange) is determined by the competition of the DMI field $H_{D}$ (violet) and the DW anisotropy field $H_{K}$ (pink). (d) The internal angle can be further tuned using an external magnetic field $H_x$ (light blue) or $H_y$ (not shown). (e) DW resistance measurements can be used to verify the DW angle $\phi$ under influence of $H_x$ and $H_y$. This can be linked to DW motion measurements on the same samples (f), where the depinning field $H_{z,\rm{depin}}$ as a function of current density $J$ is measured for the two outermost DWs. The slope of this relation defines the depinning efficiency $\epsilon$ of each DW, which scales with $\cos \phi$. The Kerr micrographs show the well-controlled alternating pattern of up (white) and down (grey) domains, realized by Ga irradiation of the white areas. }
\end{figure*}

Current-induced motion of magnetic DWs in materials with perpendicular magnetic anisotropy (PMA) could be used to transport data in next-generation storage devices \cite{Parkin2008}. Recently, it has been suggested that in addition to conventional bulk STT contributions \cite{Thiaville2005}, various current-induced torques relating to the high spin-orbit coupling in these materials could play a dominant role \cite{Miron2011a,Liu2012,Garello2013}. Most notably, the sources of these so-called spin orbit torques include the Rashba field, which enters as a current-dependent transverse $H_y$ field \cite{Miron2011,Wang2012a,Kim2012b}, and the spin Hall effect \cite{Hirsch1999}, which leads to a vertical spin current with transverse polarization $\sigma_y$. In a previous work \cite{Haazen2013}, we demonstrated that the Spin Hall effect has the correct characteristics to describe the effect of current on domain walls in Pt/Co/Pt. It was observed that the efficiency of current-induced DW motion is practically zero, since the Bloch wall that is expected to be stable does not have the correct symmetry to be moved by a SHE-induced torque (\subfigref{fig:dmi:Figure1}{a}), i.e. the cross product of the injected spin direction and magnetization direction within the DW vanishes \cite{Haazen2013,Khvalkovskiy2013}. Efficient motion arose when the internal structure was forced to the N\'eel type by applying a field along the current direction. However, this still contradicts the uniform motion of all DWs, at zero in-plane field, that was observed in other materials where the magnetic layer was sandwiched between two different materials \cite{Moore2008,Miron2011,Koyama2013}. Subsequently, it was recognized by several authors \cite{Thiaville2012,Ryu2013,Emori2013a,Emori2013b,Torrejon2013} that in the case of structural inversion asymmetry (SIA), the Dzyaloshinskii-Moriya interaction (DMI) \cite{Moriya1960} gives rise to chiral spin structures, in this case chiral N\'eel walls \cite{Heide2008,Chen2013,Chen2013a}, which are moved uniformly by a spin-Hall-effect-induced effective field ($H_{\rm{SHE}}$ in \subfigref{fig:dmi:Figure1}{b}).

In this work, we demonstrate by current-assisted depinning measurements that a measurable DMI is also present in Pt/Co/Pt \cite{Je2013}, which is surprising at first since the top and bottom interfaces are in principle the same. The DMI effect is found to be highly tunable by varying the top layer thickness, and becomes very large when the top Pt layer is substituted by AlOx. In fact, the tuning is so delicate that the DW can have any in-plane angle $\phi$ in between the Bloch ($\phi = \frac{\pi}{2}$) and N\'eel ($\phi = 0$) states at remanence, due to the competition between the effective longitudinal DMI field $H_{D}$ and transverse DW anisotropy field $H_{K}$, as visualized in \subfigref{fig:dmi:Figure1}{c}. Our data can be explained by a very simple model \cite{Thiaville2012} of the internal DW angle $\phi$ under influence of $H_{D}$, $H_{K}$, and externally applied in-plane fields $H_x$ and $H_y$ (\subfigref{fig:dmi:Figure1}{d}). The efficiency of DW depinning is simply proportional to $\cos \phi$, as expected from a field-like torque by the SHE \cite{Thiaville2012}. To prove that the DW angle $\phi$ is responsible for the efficiency, we measure the DW resistance as a function of in-plane fields (\subfigref{fig:dmi:Figure1}{e}) and observe that the DW structure is indeed changing from Bloch to N\'eel, which is often just assumed based on elementary micromagnetics without any convincing experimental proof. These measurements allow us to directly correlate a high DW efficiency to the presence of N\'eel walls, which we show to be stabilized by the DMI in a tunable way.

To allow for DW resistance as well as DW depinning measurements, 1.5\,$\mu$m wide strips with varying layer configurations were fabricated. Irradiation with Ga ions is employed to locally reduce the PMA in 1.5$\mu$m long areas in these strips, allowing us to introduce a well-defined number of DWs into the strip \cite{Franken2011,Franken2012b}. The Kerr microscopy image in \subfigref{fig:dmi:Figure1}{e} visualizes the controlled domain structure that is essential for the DW resistance measurements presented later. When the perpendicular field strength is increased from this state, DWs are randomly depinning from the edges of the irradiated regions, as indicated in \subfigref{fig:dmi:Figure1}{f}. We analyze the effect of current on the depinning of two particular DWs, indicated by the red circle and blue square in \subfigref{fig:dmi:Figure1}{f}. The graph shows how the depinning field of these domain walls change with increasing current density in a Pt(4)/Co(0.4)/Pt(2) sample (all thicknesses in nm). The slopes define the depinning efficiency $\epsilon = \mu_0 \frac{\mathrm{d} H_{\rm{SHE}} }{ \mathrm{d} J}$ of each DW. The opposing slopes of the 2 DWs actually imply a small but uniform action of the current on the DWs: since the field pushes both DWs outwards (in opposite directions), the current reduces the depinning field of DW2 (negative efficiency) and increases the depinning field of DW1 (positive efficiency). It is worth noting that the sign of current-induced domain wall motion opposes the electron flow direction and is therefore unlikely to be caused by conventional STT. Instead, we propose that the DWs have a small degree of built-in chirality, which leads to their uniform motion driven by the SHE. This is a refinement to our observations on similar samples in \cite{Haazen2013}, where we assumed that domain-walls are of the non-chiral Bloch type at zero in-plane field. To prove the presence of a favored chirality in Pt/Co/Pt, we will use in-plane fields to either oppose or assist the built-in chiral field $H_{D}$.

\begin{figure*}[tbhp]
\includegraphics[width=\linewidth]{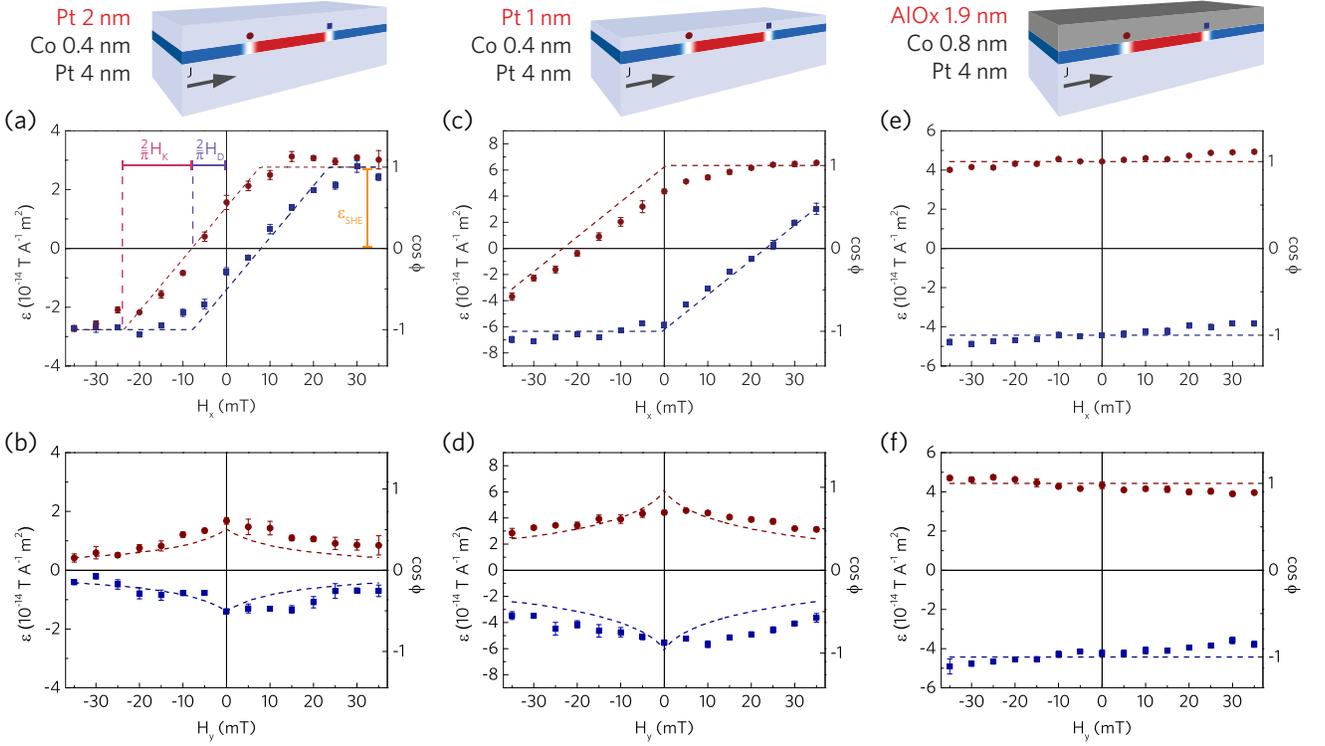}
\caption{\label{fig:dmi:Figure2} Depinning efficiency $\epsilon$ under in-plane fields in systems with varying SIA. (a) in almost symmetric Pt(4)/Co(0.4)/Pt(2), $\epsilon(H_x)$ shows a small opening between the two domain walls (red discs and blue squares) of opposite polarity, indicative of a DMI field $H_{D}$. The slope of the transition from  $\epsilon = 0$ to its saturation level $\epsilon_{\rm{SHE}}$ is characterized by the DW anisotropy $H_K$. To determine these parameters, a global fit (dashed lines) is performed together with the $H_y$-data in (b). (c,d) Increasing the stack asymmetry by reducing the top Pt layer to 1\,nm gives a higher DMI opening. (e,f) Replacing the top Pt layer by AlOx, the DMI opening becomes too large to measure in our setup. The DMI is so strong that the DW angle $\phi$ is hardly influenced by $H_x$ and $H_y$. The fit corresponds to the average level.}
\end{figure*}

\figref{fig:dmi:Figure2} shows the measured current-induced depinning efficiency $\epsilon (H_x)$  (top panes) and $\epsilon (H_y)$ (bottom panes) on three samples with different compositions. We first discuss \subfigref{fig:dmi:Figure2}{a-b}, representing the sample with the lowest degree of inversion asymmetry, Pt(4)/Co(0.4)/Pt(2). The $H_x$ and $H_y$ data have been fitted simultaneously using the efficiency expected from the 1D-model with only the SHE as driving force \cite{Thiaville2012},
\begin{equation}\label{eq:dmi:efficiency}
\epsilon = \mu_0 \frac{\mathrm{d} H_{\rm{SHE}}} {\mathrm{d} J} = \frac{\pi \hbar \nu \theta_{\rm{SH}}}{4 e M_{\rm{s}} t} \cos \phi =: \epsilon_{\rm{SHE}}   \cos \phi ,
\end{equation}
with $\theta_{\rm{SH}}$ the spin Hall angle, $M_{\rm{s}}$ the saturation magnetization, $t$ the magnetic layer thickness, $\nu$ a loss factor due to compensating spin Hall currents from the bottom and top Pt layers \cite{Haazen2013}, and $\phi$ the internal DW angle prescribed by energy minimization of
 \begin{eqnarray}\label{eq:dmi:angle}
\nonumber E_{\rm{DW}} = \lambda \mu_0 M_{\rm{s}} \left(  H_{K} \cos^2 \phi \phantom{\frac{\pi}{2} \frac{\pi}{2} \frac{\pi}{2}\frac{\pi}{2}\frac{\pi}{2}\frac{\pi}{2}} \right. \\ \left. -2 \left(H_{D} + \frac{\pi}{2} H_x \right) \cos \phi  - \pi H_y  \sin \phi \right).
\end{eqnarray}
Here, $\lambda$ is the DW width (assumed constant), $H_K$ the shape anisotropy field of the DW, and $H_D = \pi D / (2 \mu_0 M_{\rm{s}} \lambda)$ the effective chiral magnetic field (with $D$ an energy constant characterizing the strength of the DMI). The free parameters of the fit are $H_K$, $H_D$, and $\epsilon_{\rm{SHE}}$ as indicated in \subfigref{fig:dmi:Figure2}{a}. This graph clearly shows that there is a contribution from DMI: the red and blue curves have been shifted to the left and right, respectively, due to the effective chiral magnetic field $H_D\approx 12.5 \pm 0.4\,$mT, which has opposite sign for domain walls of opposite polarity (up-down vs down-up). We should note that $H_D$ could be lower than in an unpatterned film, since the Ga irradiation locally reduces the anisotropy and increases $\lambda$. Apart from the observed horizontal shift, there is a linear increase from $\epsilon = 0$  to $\epsilon_{\rm{SHE}}$ over a field range $\pi H_{K} /2$, which is attributed to the transition from Bloch to N\'eel. Replacing $H_x$ by $H_y$ in \subfigref{fig:dmi:Figure2}{b}, the efficiency simply decreases with $|H_y|$, because $H_y$ gradually pulls the wall to a Bloch state. Interestingly, the DW at zero in-plane field is neither a Bloch nor a N\'eel wall. From the efficiency at zero in-plane field, it can be deduced that the DW angle at remanence is $\phi \approx 60^{\circ}$, rather than the $90^{\circ}$ that is expected in a system without SIA. The stability of such an in-between wall type, observed before on epitaxial Co/Ni multilayers \cite{Chen2013a}, might be interesting for specific device applications of sputtered PMA films.

\begin{figure*}[tbhp]
\includegraphics[width=\linewidth]{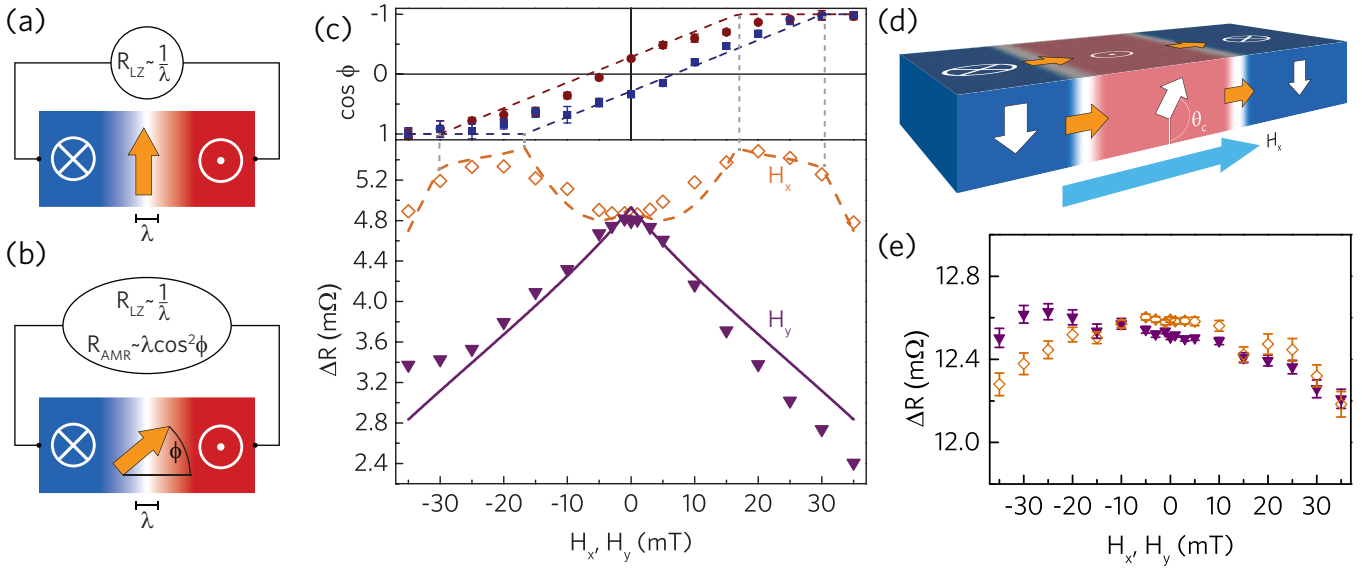}
\caption{\label{fig:dmi:Figure3} Bloch-N\'eel transition revealed by DW resistance measurements. (a) a Bloch wall yields only an intrinsic contribution $R_{\rm{LZ}}$ to the DW resistance. (b) As $\phi$ approaches 0 (N\'eel wall), an additional contribution from the AMR effect arises. (c) (top) Depinning efficiency as a function of $H_x$ and (bottom) Resistance change induced by 20 DWs of alternating polarity, as a function of $H_x$ (open diamonds) and $H_y$ (filled triangles), measured on the same Pt(4)/Co(0.5)/Pt(2) sample. The dashed orange and solid purple line are a fit including the two resistance contributions, in which $H_K$ and $H_D$ have been taken from the fit of the depinning data. The DW resistance peaks around $H_x=20\,$mT due to the transition from Bloch to N\'eel. A decreasing background signal is present in both the $H_y$ and the $H_x$ data due to canting of the magnetization in the irradiated domain (d), leading to a lower intrinsic contribution $R_{\rm{LZ}}$. (e) DW resistance in Pt/Co/AlOx does not show the Bloch-N\'eel transition since N\'eel walls are highly stable. }
\end{figure*}

To explore the tunability of the small DMI in Pt/Co/Pt, samples with a thinner 1\,nm Pt capping layer were fabricated, in order to increase the apparent SIA. Indeed, in \subfigref{fig:dmi:Figure2}{c} it is observed that $H_D$ has increased significantly to the value $37\pm1\,$mT. The change of efficiency $\varepsilon_{\rm{SHE}}$ matches with the change in layer thickness as discussed in the Supplementary Information. The SIA can be increased much more by replacing the top Pt layer by a different material, AlOx, as shown in \subfigref{fig:dmi:Figure2}{e-f}. In fact, $H_D$ has become so large that we cannot quantify it within our setup. Regardless of the in-plane field, the domain-walls are chiral N\'eel walls which are pushed uniformly in the direction of current flow, hence opposite to conventional STT. A small linear effect of the in-plane field on the efficiencies appears to be present, which is either an experimental artifact or caused by mechanisms beyond our simplified 1D model. We verified this effect does not have the correct characteristics to be described by the Rashba effect, by quasi-statically reproducing the depinning process in the dynamic 1D DW model \cite{Thiaville2012} including a current-dependent $H_y$ Rashba field. Crucially, these results suggest that our Pt/Co/AlOx is not fundamentally different from Pt/Co/Pt, but only has a higher built-in chiral field due to the increased asymmetry.

Although the results presented so far here and by other authors \cite{Thiaville2012,Ryu2013,Emori2013a,Emori2013b,Torrejon2013} match well with a SHE-induced torque dictated by the DW angle $\phi$, it is not at all trivial that the DW indeed has the structure that these experiments suggest. In fact, the transition from a Bloch-like to a N\'eel state by an in-plane field was to our knowledge not measured before. We believe that using DW resistance as a probe for the DW structure is much more direct than using the DW mobility, since the resistance does not rely on any of the spin torques. Therefore, we have performed measurements of the DW resistance as a function of in-plane field. To be able to accurately measure tiny resistance changes induced by the DWs, we use lock-in measurements on an on-sample Wheatstone bridge\cite{Aziz2006,Franken2012b} consisting of four nominally identical wires, one of which has a Ga-irradiation pattern. To exclude magnetoresistive effects in the domains from polluting the measured resistance of the DWs, we measure the bias of the bridge for the monodomain state at each in-plane field, and subtract it from the bias in the presence of the multidomain state shown in \subfigref{fig:dmi:Figure1}{e} (see also Methods section). When the resistance of a Bloch wall is measured (\subfigref{fig:dmi:Figure3}{a}), the dominant contribution comes from the intrinsic resistance $R_{\rm{LZ}} \sim 1/\lambda$ dictated by the Levy-Zhang model \cite{Levy1997} for which we recently found experimental evidence \cite{Aziz2006,Franken2012b}. However, when the magnetization within the DW obtains a component parallel to the current flow (\subfigref{fig:dmi:Figure3}{b}), an additional contribution $R_{\rm{AMR}}$ from anisotropic magnetoresistance arises \cite{Koyama2011}, simply proportional to $\lambda \cos^2 \phi$.

Looking at the measured DW resistance in Pt(4)/Co(0.5)/Pt(2) as a function of $H_x$ (open orange diamonds in \subfigref{fig:dmi:Figure3}{c}), we indeed see an increase when the in-plane field increases, owing to the transformation to a N\'eel wall. However, beyond $H_x=30\,$mT, the measured DW resistance starts to decrease again. This decrease is seen over the entire $H_y$-field range (purple triangles), which can be regarded as a kind of background measurement. We attribute this to a change of $R_{\rm{LZ}}$ related to the domain structure. Since the anisotropy in one of the domains is reduced strongly by the Ga irradiation treatment \cite{Franken2012}, this region tends to be pulled in plane, modifying the domain-wall profile as sketched in (\subfigref{fig:dmi:Figure3}{d}). This has a strong effect on $R_{\rm{LZ}}$, and also a small effect on $R_{\rm{AMR}}$  (see Supplementary Information for details of the model). From the DW-depinning data (top pane of \subfigref{fig:dmi:Figure3}{c}), we can deduce $\phi(H_x, H_y)$, and use this as input in our resistance model. The best-fit to this model is presented as the dashed orange and solid purple lines in \subfigref{fig:dmi:Figure3}{c}), where the free parameters are the anisotropy in the irradiated domain, the AMR resistivity, and the strength of the intrinsic DWR. The model reproduces the measurements, apart from two kinks at each field polarity (when the angles of either the 'red' or the 'blue' domain walls saturate, compare to top panel). It is not surprising that these sharp features from the 1D model become smooth in reality, especially since we measure the sum of 20 DWs, each with slightly different local properties.

Looking at the DW resistance measurements in Pt/Co/AlOx in \subfigref{fig:dmi:Figure3}{e}), only minor changes as a function of in-plane field are observed, and there is no clear difference between the $H_x$ and $H_y$ data. Similar to the DW-depinning results, this suggests that the N\'eel character of the DWs is stabilized by a strong built-in chiral field and is not significantly influenced by the range of applied in-plane fields. The background due to magnetization canting is largely absent here, because the anisotropy in the irradiated domain is much higher. Although any change as a function of in-plane field is relatively small compared to the Pt(4)/Co(0.5)/Pt(2) sample, a small asymmetry appears in the $H_y$ data, which may have the same unknown origin as the slight asymmetry in the DW depinning data in \subfigref{fig:dmi:Figure2}{f}.

We should note that, in our analysis so far, we assumed that the DW is oriented perpendicular to the nanowire, whereas it was recently demonstrated that the DW boundary might tilt in the $xy$-plane in systems with a significant DMI \cite{Emori2013b,Boulle2013}. Such tilting will occur when the DW moves at high speed (which is not applicable in depinning measurements), or when $H_y$ fields are applied \cite{Boulle2013}. We estimate that the highest applied $H_y$ of $\sim 40$\,mT, could induce a tilting of at most $10^{\circ}$ in Pt/Co/AlOx, which has the highest DMI. We do not observe significant tilting in the Kerr images (within the $\sim 300$\,nm resolution), nor do we observe a dramatic difference between the DW resistance as a function of $H_y$ and $H_x$ in this material, hence the influence of this possible tilt angle is limited. Furthermore, in Pt/Co/Pt samples the DMI is an order of magnitude smaller, hence a tilting of at most 1 or 2 degrees might be induced, which is hardly significant and therefore not taken into account in the analysis.

We now briefly discuss the origin of the unexpected chiral effective fields in Pt/Co/Pt. It was recently calculated \cite{Freimuth2013} that a significant DMI can arise in a Pt/Co bilayer, leading to a effective chiral field of several 100\,mT. Since the chiral field in Pt/Co/Pt is the result of two canceling interfaces, we should stress that an imbalance between the DMI at the top and bottom interface of only a few percent is enough to achieve the measured magnitude. Ryu et al.~\cite{Ryu2013} also studied the effect of stack asymmetry on the effective chiral field in Co/Ni multilayers, and concluded that the DMI originates at Pt/Co interfaces and scales with the thickness of the neighboring Co layer, which they attributed to proximity-induced moments in Pt. They concluded that DW motion in the direction of current flow implies that the DMI at the bottom interface dominates over the top interface. In the current manuscript, we appear to have tuned the DMI through the thickness of the Pt layers themselves. Due to growth-related phenomena, it is known that the top and bottom interface can have different characteristics, for example evidenced by a different contribution to the effective PMA \cite{Bandiera2011}. Apparently, the DMI at the top interface decreases when reducing the top layer thickness, such that the net DMI increases. It is worth noting that although the DMI at the top layer appears to decrease, the PMA constant increases for thinner Pt top layers (see Supplementary Information Table 1). In an inverted Pt(2)/Co/(0.5)/Pt(4) sample (Supplementary Information, Figure S1), it was found that the DMI almost vanishes, but does not change sign, thus the interfaces appear to become more symmetric for thick capping layers. This indeed suggests that interface characteristics are key, rather than the thickness of the layers themselves. Reasons for the top interface to vary with thickness can be changes to the mode of growth, different interdiffusion \cite{Bandiera2011} or even slight oxidation at the top Co/Pt interface in case of thin capping layers.

To summarize, the effective chiral magnetic field in Pt/Co systems turns out to be tunable by varying the top layer thickness and material. The effect scales with the degree of structural inversion asymmetry and leads to a gradual change of the stable wall type from Bloch to N\'eel. Furthermore, by using the DW resistance as an independent measurement of the internal DW structure, a change of the internal structure from Bloch-like to N\'eel under longitudinal fields was evidenced, and correlated to the high efficiency of DW motion of N\'eel walls. These findings firmly establish SHE and DMI as a tandem for efficient and uniform domain wall motion.

\section{Methods}
\subsection{Sample fabrication}
All samples consisted of 1.5\,$\mu$m wide strips fabricated on Si/SiO$_2$ substrates by Electron-Beam lithography, DC sputtering, and lift-off. The Pt(4)/Co(0.8)/Al(1.5) samples were oxidized in a 15\,W, 0.1\,mbar O$_2$ plasma for 10 minutes to obtain Pt/Co/AlOx. These samples were annealed for 20 minutes at 573\,K. The Pt/Co/Pt samples did not undergo an annealing treatment. The samples were designed to form an on-sample Wheatstone bridge configuration to be able to measure resistance changes accurately (for details see \cite{Franken2012b}). The samples were locally irradiated with a 30keV Ga FIB to make it possible to create a stable domain pattern with a well-defined number of domain walls. At the same time, the edges of the irradiation boundaries acted as pinning sites, to enable the well-controlled depinning measurements.  The Ga doses were chosen to ensure that all DWs are stable during DWR measurements and amounted to $1.13\times10^{13}\,$ions/cm$^2$ on Pt/Co/AlOx, $0.50\times10^{13}\,$ions/cm$^2$ on Pt(4)/Co(0.5)/Pt(2) and Pt(4)/Co(0.4)/Pt(2), and $0.81\times10^{13}\,$ions/cm$^2$ on Pt(4)/Co(0.4)/Pt(1). Table S1 (Supplementary Information) provides the material parameters (PMA constant and $M_{\rm{s}}$) obtained by VSM-SQUID magnetometry on unpatterned films.

\subsection{DW depinning measurements}
Very low DC current densities in the range $\pm2\times10^{10}$\,A/m$^2$ were used to exclude significant effects from Joule heating and Oersted fields. The current and in-plane field are kept constant, while the $z$-field is being ramped up until both DWs have depinned which is automatically detected by an image analysis routine. This is repeated at least 10 times for each current point to obtain sufficient signal to noise. Table S1 presents the fit parameters of the 1D model ($H_K, H_D$ and $\epsilon_{\rm{SHE}}$) for each used material composition, and compares the latter to the expected efficiency  $\epsilon_{\rm{calc}}$ based on the layer thicknesses.

\subsection{DW resistance measurements}
Our measurement routine is very similar to what we described in \cite{Franken2012b}. We use a combination of an on-sample Wheatstone bridge and a lock-in technique to measure the resistance change due to the presence of domain walls. The in-plane field is applied constantly, and first the wire is saturated (zero DWs) by a negative $H_z$, and the lock-in voltage at $H_z \approx -1\,$mT is recorded. Then, the domain walls are created by a positive $H_z$, the field is reduced to $H_z \approx 1\,$mT and the lock-in voltage is recorded again. The difference is presented in \figref{fig:dmi:Figure3} as $\Delta R$, and represents the resistance change due to all the DWs (20 or 18 for Pt/Co/Pt and Pt/Co/AlOx, respectively). Since the background voltage is recorded at the same in-plane field as the voltage in the presence of DWs, magnetoresistive effects within the bulk of the domains, such as AMR due to canting or the magnon contribution \cite{Nguyen2011a}, are automatically filtered out, leaving only resistance changes in the DW region.We always check the number and positions of the DWs present by real-time comparison to a Kerr-microscopy image. Furthermore, we always ensure that the magnetization underneath the 20\,nm thick Pt contacts and the three reference strips in the bridge does not switch, as this can give rise to additional magnetoresistive signals. The small $H_z$ during measurements serves to ensure the DW pattern remains stable during measurements. An AC probe current of 1\,mA (0.75\, mA) at 500\,Hz was sent through two parallel series of Pt(4)/Co(0.5)/Pt(2) (Pt(4)/Co(0.8)/AlOx) wires and it was verified that a lower amplitude does not significantly alter the results. Because the bridge is not perfectly balanced, a bias of typically 10\,mV occurs, even when all strips are magnetized in the same direction. When we introduce domain walls into one of the strips, the signal typically changes by 10\,$\mu$V. Note that a sample with slightly thicker $t_{\rm{Co}}=0.5\,$nm had to be used, because the samples with  $t_{\rm{Co}}=0.4\,$ turned out to be very easily switched by the spin Hall effect from the probe current. Therefore, much lower probe currents have to be used, and the thinner magnetic Co layer further deteriorates the signal/noise. Details on the modeling of in-plane field effects on the measured DW resistance are presented in the Supplementary Information, and the individual contributions to the modeled resistance are plotted in Figure S2.

\newpage 
\section*{Acknowledgements}  
This work is part of the research programme of the Foundation for Fundamental Research on Matter (FOM), which is part of the Netherlands Organisation for Scientific Research (NWO).

\section*{Author Contributions}
J.H.F. designed the experiments and prepared the manuscript. M.H. and J.H.F. performed the experiments and the data analysis.  H.J.M.S. and B.K. assisted in the analysis and commented on the final manuscript.

\newpage
\pagebreak
\widetext
\begin{center}
\textbf{\large Supplementary Information: Tunable chiral spin texture in magnetic domain-walls}
\end{center}
%%%%%%%%%% Merge with supplemental materials %%%%%%%%%%
%%%%%%%%%% Prefix a "S" to all equations, figures, tables and reset the counter %%%%%%%%%%
\setcounter{equation}{0}
\setcounter{section}{0}
\setcounter{figure}{0}
\setcounter{table}{0}
\setcounter{page}{1}

\renewcommand*{\citenumfont}[1]{S#1}
\renewcommand*{\bibnumfmt}[1]{[S#1]}

\renewcommand{\figurename}{Figure}
\renewcommand{\tablename}{Table}
\renewcommand{\thetable}{S\arabic{table}}
\renewcommand{\thefigure}{S\arabic{figure}}
\renewcommand{\theequation}{S\arabic{equation}}
\renewcommand{\thesection}{\arabic{equation}}

% Usual (decimal) numbering
\renewcommand{\thesection}{S\arabic{section}}
\renewcommand{\thesubsection}{\thesection.\arabic{subsection}}
\renewcommand{\thesubsubsection}{\thesubsection.\arabic{subsubsection}}
% Fix references
\makeatletter
\renewcommand{\p@subsection}{}
\renewcommand{\p@subsubsection}{}
\makeatother

\section{Material parameters}

\begin{table}[h]
\begin{tabular}{|l|c|c|c|c|c|c|c|}
\hline
   & $M_{\rm{s}}$ & $K_{\rm{eff}}$ & $\mu_0 H_D$ & $\mu_0 H_K$ & $\epsilon_{\rm{SHE}}$   & $\epsilon_{\rm{calc}}$  & $\nu$ \\
   & MA/m         & MJ/m$^{3}$     & mT          & mT          & \multicolumn{2}{c|}{$10^{-14}$\,TA$^{-1}$m$^{2}$} &       \\ \hline
Pt(4)/Co(0.36)/Pt(1)    & 1.01(5)      & 0.39(2)        & 37(1)       & 37(1)       & 6.4(2)                  & 6.7                     & 0.68  \\
Pt(4)/Co(0.36)/Pt(2)    & 1.08(5)      & 0.27(1)        & 12.5(4)     & 24.6(5)     & 2.76(6)                 & 3.2                     & 0.34  \\
Pt(4)/Co(0.8)/AlOx(1.9) & 1.17(5)      & 0.28(1)        & $\gg 40$           & ?           & 4.43(6)                 & 3.4                     & 0.89  \\
Pt(4)/Co(0.5)/Pt(2)     & 1.07(5)      & 0.28(1)        & 11(2)       & 37(1)       & 2.4(1)                  & 2.3                     & 0.34  \\
Pt(2)/Co(0.5)/Pt(4)     & 1.18(5)      & 0.22(1)        & 3(1)        & 19(2)       & -1.8(1)                 & -2.1                    & -0.34 \\ \hline
\end{tabular}
\caption{ \label{table:dmi_supp:properties} Fit parameters and material properties of various compositions.  }
\end{table}

\begin{figure}[h]
\includegraphics[width= 0.5\linewidth]{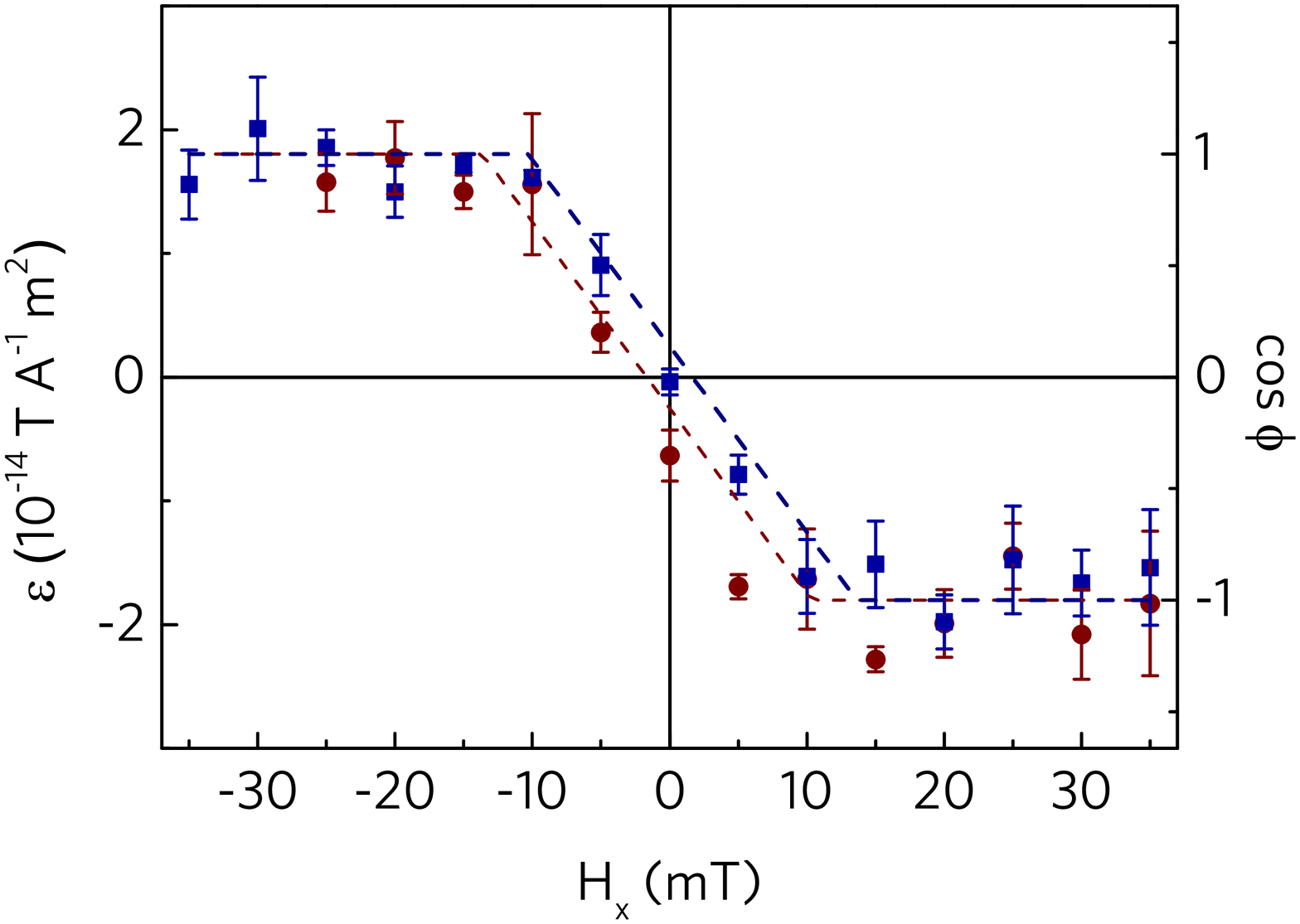} \caption{\label{fig:dmi_supp:Pt2Co0.5Pt4} Depinning efficiency as a function of $H_x$ on the inverted stack Pt(2)/Co(0.5)/Pt(4). }
\end{figure}

Table \ref{table:dmi_supp:properties} summarizes the measured material properties. $M_{\rm{s}}$ and $K_{\rm{eff}}$ have been measured by VSM-SQUID magnetometery of unpatterned films. The samples labeled with a Co thickness of 0.4\,nm in the text for convenience, were actually 0.36\,nm thick. $\mu_0 H_D$, $\mu_0 H_K$, and $\epsilon_{\rm{SHE}}$ have been obtained by fits of the DW depinning data like in Figure 2 in the main text. Although a stronger spin Hall current is injected in Pt/Co/AlOx, $\epsilon_{\rm{SHE}}$ is smaller than in Pt/Co/Pt(1\,nm) because it is absorbed by a thicker Co layer (see equation (1) in the main text).  The inverted stack Pt(2)/Co(0.5)/Pt(4) also has an inverted $\epsilon_{\rm{SHE}}$, as we already explained in \cite{Haazen2013}. \figref{fig:dmi_supp:Pt2Co0.5Pt4} provides a new measurement of the depinning efficiency as a function of $H_x$ on this layer system, in order to reveal the presence of DMI. There is a very small opening visible, indicative of a $H_{D}$ with the same sign as the inverted composition. Actually, it seems like one of the DWs has zero $H_{D}$ (crosses through the origin), whereas the other one has a small but finite $H_{D}$. In any case, this suggests that the DMI, unlike the SHE, is not a result of the Pt layer thicknesses themselves, but rather the effect of increasing asymmetry between the top and bottom interface when the top layer is varied.

Since both DMI and PMA are expectedly interface effects, it is interesting to look for correlations between the parameters $H_{D}$ and $K_{\rm{eff}}$. For the Pt/Co/Pt samples, there is indeed a positive correlation between $H_D$ and $K_{\rm{eff}}$. However, Pt/Co/AlOx breaks this trend: it has a much stronger $H_{D}$ than any other sample whereas $K_{\rm{eff}}$ is similar, so the two parameters are definitely not always directly related. Given that the DW motion is in the direction of current flow, we know that the DMI at the bottom interface must be dominant over the DMI from the top interface \cite{Ryu2013}. So in fact, the DMI at the top interface must \emph{decrease} when the top layer is made thinner, whereas the anisotropy contribution from this interface is actually seen to \emph{increase}. So there appears to be a \emph{negative} correlation between the anisotropy and DMI at the top interface, leading to a \emph{positive} correlation between the anisotropy and the total DMI which is dominated by the bottom Pt/Co interface.

\subsection{Spin Hall amplitudes}
In the last two columns of Table \ref{table:dmi_supp:properties}, we have calculated the expected loss factor $\nu$ of the spin Hall effect, and the accompanying depinning efficiency of N\'eel walls in the 1D model \cite{Thiaville2012},
\begin{equation}\label{eq:dmi_supp:efficiencyCalc}
\epsilon_{\rm{calc}} = \frac{\pi \hbar \nu \theta_{\rm{SH}}}{4 e M_{\rm{s}} t}.
\end{equation}
The calculation of $\nu$ is straightforward; the net spin Hall current due to a \emph{single} thin Pt layer with thickness $t_{\rm{Pt}}$  is given by \cite{Liu2011a}
\begin{equation}
J_{\rm{S}} (t_{\rm{Pt}})  = \theta_{\rm{SH}} J \left( 1- \mathrm{sech}\left(\frac{t_{\rm{Pt}}}{\lambda_{\rm{sf}}}\right)\right),
\end{equation}
where $\lambda_{\rm{sf}} \approx 1.4\,$nm the spin diffusion length of Pt \cite{Liu} and $\theta_{\rm{SH}} = 0.07$ the spin Hall angle of Pt \cite{Liu}. For a Co layer sandwiched between two Pt layers, two of these spin currents with opposite polarization are injected, yielding a net spin current
\begin{equation}
J_{\rm{S}}^{\rm{eff}} = \nu \theta_{\rm{SH}} J := \left( \mathrm{sech}\left(\frac{t^{\rm{top}}_{\rm{Pt}}}{\lambda_{\rm{sf}}}\right)- \mathrm{sech}\left(\frac{t^{\rm{bottom}}_{\rm{Pt}}}{\lambda_{\rm{sf}}}\right) \right) \theta_{\rm{SH}} J.
\end{equation}

Comparing the calculated $\epsilon_{\rm{calc}}$ to the measured $\epsilon_{\rm{SHE}}$ in Table \ref{table:dmi_supp:properties}, we observe close agreement. The largest deviation is found in Pt/Co/AlOx which measures a slightly higher $\epsilon_{\rm{SHE}}$ than expected. We should note that this is the only sample that has undergone an annealing treatment, hence it might have different properties compared to the other ones. There might also be a contribution from conventional STT to $\epsilon_{\rm{SHE}}$ in Pt/Co/AlOx, but since conventional STT would oppose the SHE torque, this should reduce the measured $\epsilon_{\rm{SHE}}$ compared to the model, whereas the difference we observe is opposite.
%\begin{equation}
%J_{\rm{S}}^{eff} := \nu \theta_{\rm{SH}} J = \left\{  \left( 1- \mathrm{sech}\left(\frac{t^{bottom}_{\rm{Pt}}}{\lambda_{\rm{sf}}}\right) \right) \frac{J_{\rm{bottom}}}{J} - \left( 1- \mathrm{sech}\left(\frac{t^{top}_{\rm{Pt}}}{\lambda_{\rm{sf}}}\right) \right) \frac{J_{\rm{top}}}{J} \right\} \theta_{\rm{SH}} J
%\end{equation}

\section{DW resistance model}
In this section, we propose a model to describe the two dominant contributions to the DW resistance. We first apply the Levy-Zhang model of the intrinsic resistivity to the expected DW profile in the sample. Then, we discuss the contribution from anisotropic magnetoresistance (AMR). Finally, an expression is given for the measured resistance change in an actual Pt/Co/Pt layer, where current shunts through the Pt layers. This expression is fitted to the experimental data.

\subsection{Levy-Zhang model for arbitrary DW profiles}
The Levy-Zhang model describes the contribution to DW resistance due to spin mistracking \cite{Levy1997}. In their original derivation, they assume a simplified DW profile of the form $\theta(x) = \pi x /d$. However, in reality the DW has the more complicated Bloch profile, and when magnetization canting due to an in-plane field starts to play a role, the actual profile is even more complex. We therefore first derive an expression valid for any DW profile, and then insert an approximated `canted' profile to find an expression for $R_{\rm{LZ}}$ as a function of in-plane field.

The original result of DW resistivity by Levy and Zhang for a current perpendicular to the DW reads
\begin{equation} \label{eq:dmi_supp:LZ}
\rho_{\rm{LZ}} = C \left(\frac{\pi}{\lambda}\right)^2,
\end{equation}
with $\lambda$ the DW width and $C$ a prefactor given by
\begin{equation} \label{eqs:dmi_supp:prefactor}
C = \frac{\hbar^4 k^2 \rho_0}{80 J^2 m^2} \left(\frac{\rho_{\uparrow} }{ \rho_{\downarrow}} -2 + \frac{\rho_{\downarrow}}{ \rho_{\uparrow}}\right)\left(3+\frac{10 \sqrt{\rho_{\uparrow} / \rho_{\downarrow}}}{\rho_{\uparrow} / \rho_{\downarrow}+1}\right),
\end{equation}
with $\hbar$ Planck's constant, $k\approx 1{\AA}^{-1}$ the Fermi wavevector, $m$ the electron mass, $J\approx 0.5\,$eV the (microscopic) exchange splitting, $\rho_{\uparrow} / \rho_{\downarrow}$ the spin asymmetry in the Co layer, and $\rho_0$ the resistivity of the Co layer.

 \eqref{eq:dmi_supp:LZ} was obtained for the simple DW profile with a constant slope  $\mathrm{d} \theta / \mathrm{d} x = \pi / \lambda$. For a real DW in which this slope is not constant, the resistivity is position-dependent within the DW. Therefore, a more general form of the DW resistivity is
 \begin{equation}
\rho_{\rm{LZ}} = C \left(\frac{\mathrm{d} \theta(x)}{\mathrm{d} x}\right)^2.
\end{equation}

The DW resistance is found by integrating the resistivity over the entire DW profile,
 \begin{equation} \label{eq:dmi_supp:RLZ}
 R_{\rm{LZ}}(x) = \frac{1}{S} \int_{-\infty }^{\infty } C \left(\frac{\mathrm{d} \theta(x)}{\mathrm{d} x}\right)^2 \, \mathrm{d}x,
\end{equation}
with $S$ the cross-sectional area of the magnetic layer.

As explained briefly in the main text, we expect at high in-plane fields a DW profile that rotates from $\theta = 0$ in the non-irradiated region, to $\theta_{\rm{c}} (H_x) < \pi$ in the Ga-irradiated region, where the anisotropy has decreased so much that the magnetization is significantly pulled in-plane. We assume a scaled Bloch profile that takes into account this smaller final angle of the DW,
\begin{equation} \label{eq:dmi_supp:profile}
\theta(x) =  \frac{2}{\pi } \theta_{\rm{c}}(H_x) \arctan \left(e^{x/\lambda}\right),
\end{equation}
where we use $\lambda = \sqrt{A/K_{\rm{low}}}$ with $A=16\,$pJ/m and $K_{\rm{low}}$ the effective anisotropy in the irradiated region (a fit parameter), which is seen to determine the DW width in micromagnetic simulations. From the Stoner-Wohlfarth model, it is straightforward to derive that the magnetization canting as a function of in-plane field is given by
\begin{equation}  \label{eq:dmi_supp:theta_c}
\theta_{\rm{c}}(H_x)=\pi -\arcsin \left(\frac{H_x M_{\rm{s}}}{2 K_{\rm{low}}}\right).
\end{equation}
Plugging the DW profile of \eqref{eq:dmi_supp:profile} into the expression for the resistance \eqref{eq:dmi_supp:profile} yields
\begin{equation}  \label{eq:dmi_supp:RLZcanted}
 R_{\rm{LZ}} (H_x) = \frac{2 C}{S} \frac{ \left(\pi -\arcsin\left(\frac{H_x M_{\rm{s}}}{2 K_{\rm{low}}}\right)\right)^2}{\pi ^2 \lambda}.
\end{equation}
Note that we used $H_x$ in the expressions above, but the same expressions hold for $H_y$.

\subsection{AMR contribution}

The AMR resistivity within the DW scales with the square of the projection of the magnetization on the $x$-axis, hence
\begin{equation} \label{eq:dmi_supp:dwamr}
\rho_{\rm{DWAMR}}(x) = \rho_{\rm{AMR}}  \cos^2\phi \sin^2 \theta(x),
\end{equation}
where $\rho_{\rm{AMR}}$ is the AMR resistivity parameter of Co. We will assume that the angle $\phi$ does not vary within the DW (which is supported by micromagnetic simulations). $\phi =0$ represents a N\'eel wall, giving the highest AMR.

If there is no DW present and no canting of the magnetization, the additional AMR contribution in the presence of DWs is found by integrating \eqref{eq:dmi_supp:dwamr}. However, if one of the domains is canted in the $x$-direction, there is a large contribution form this domain to the AMR. This is however not the experimental situation, because the subtracted background signal is recorded at the same in-plane field, hence AMR from the domains is not included in the presented DW resistance. Since we do not have an analytical expression for this background, we start from the original Bloch profile which rotates from 0 to $\pi$ so that the integral to infinity converges, and multiply \eqref{eq:dmi_supp:dwamr} by a correction factor $\cos^2\theta_{\rm{c}}(H_x)$ which is not analytical but at least correct in the center of the domain wall,
\begin{equation} \label{eq:dmi_supp:dwamrcorrect}
\rho_{\rm{DWAMR}}(x) = \rho_{\rm{AMR}}  \cos^2\phi \sin^2 \theta(x) \cos^2\theta_{\rm{c}}(H_x).
\end{equation}

Now we only need to integrate the resistivity to get the AMR contribution to the DW resistance,
\begin{equation}
R_{\rm{AMR}} = \frac{1}{S} \int_{-\infty }^{\infty } \rho_{\rm{DWAMR}} \mathrm{d} x= \frac{1}{S} \frac{\rho_{\rm{AMR}}\lambda \left(4 K_{\rm{low}}^2-H_x^2 M_{\rm{s}}^2\right) \cos^2\phi}{2 K_{\rm{low}}^2}.
\end{equation}

\subsection{Converting to actually measured resistance change}

The actually measured resistance change is reduced strongly by current shunting through the Pt layers. Assuming only a fraction $p\approx 0.03$ of the current runs through the Co layer in Pt/Co/Pt based on a Fuch-Sondheimer model \cite{Franken2012b}, the resistance of the wire $R_{\rm{wire}}$ can be described as the result of two parallel resistors $R_{\rm{Co}} = \frac{R_{\rm{wire}}}{p}$ and $R_{\rm{Pt}} = \frac{R_{\rm{wire}}}{1 - p}$. The occurrence of $N$ DWs only trigger a resistance change of the Co layer $\Delta R_{\rm{Co}}$,
\begin{equation}
\Delta R_{\rm{Co}} = N (R_{\rm{LZ}} + R_{\rm{AMR}}).
\end{equation}
In the parallel resistor model, it is easy to show that this leads to a resistance change of the whole wire of
\begin{equation}
\Delta R = \frac{N p^2 (R_{\rm{LZ}} + R_{\rm{AMR}}) R_{\rm{wire}}}{R_{\rm{wire}}-N (p-1) p (R_{\rm{LZ}} + R_{\rm{AMR}})}.
\end{equation}
In the Pt(4)/Co(0.5)/Pt(2) wire, $R_{\rm{wire}} = 1.3$\,k$\Omega$ and $N=20$, whereas in the Pt/Co/AlOx wire, $R_{\rm{wire}} = 1.8$\,k$\Omega$ and $N=18$

This model for $\Delta R$ has been fitted to the DWR data in Figure 3(c) in the main text, with $\rho_{\rm{AMR}}$, $C$, and $K_{\rm{low}}$ as free parameters. The value for $\rho_0$ in the prefactor $C$ was calculated as $R_{\rm{Co}}\,S/L$, with $L$ the length of the wire. Note that a dependence on the DW angle $\phi$ enters in the model via $R_{\rm{AMR}}$. The value of $\phi$ at each $H_x$ and $H_y$ are described by minimization of Eqn.~(1) in the main text, where $H_D$ and $H_K$ are extracted from the DW depinning data (see table \ref{table:dmi_supp:properties}).
The best fit was obtained with parameters $\rho_{\rm{AMR}} = 2.9\times10^{-9}\,\Omega$m, $C=2.25\times10^{-24} \,\Omega$m$^3$, $K_{\rm{low}} = 29.8\,$kJ/m$^3$. The value of the prefactor $C$ implies via \eqref{eqs:dmi_supp:prefactor} that $\rho_{\uparrow} / \rho_{\downarrow} \approx 15$, which is reasonable according to the original paper by Levy and Zhang \cite{Levy1997}. The value for $K_{\rm{low}}$ at a dose of $0.50\times10^{13}\,$ions/cm$^2$ is somewhat lower than we measured before \cite{Franken2012}, which could relate to some of the assumptions in our modeling, such as the chosen values of the fixed parameters or the assumption that the DW width does not depend on in-plane field. Note that we did not have to include additional magnetoresistance effects such as the geometric size effect or the anisotropic interface magnetoresistance \cite{Kobs2011} to obtain a reasonable fit. The presence of such an effect could alter the fit parameters, but our main conclusion that Bloch walls transform to N\'eel walls is robust simply because of the very different response to $x$ and $y$ fields, regardless of the precise relative magnitude of the effects that are responsible for the measured changes.

\begin{figure}[h]
\includegraphics[width= 0.5\linewidth]{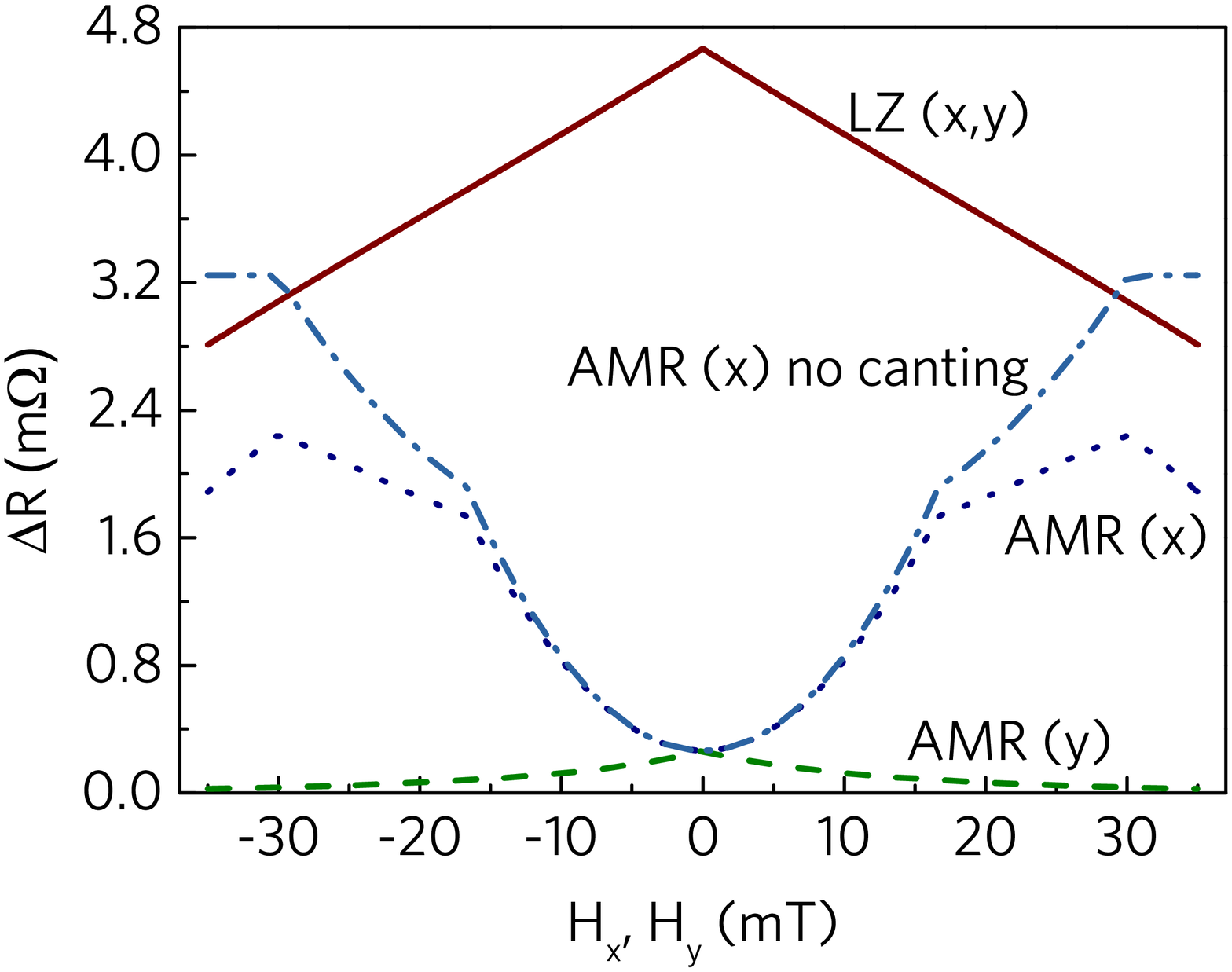} \caption{\label{fig:dmi_supp:DWR_contributions} Contributions of the various domain wall resistance effects as a function of in-plane field. Plotted are the intrinsic Levy-Zhang contribution (solid red line), the DW AMR as a function of $H_y$ (dashed green line) and $H_x$ (dotted dark-blue line), and the (hypothetical) DW AMR contribution as a function of $H_x$ in the absence of magnetization canting (dash-dotted light-blue line). The kinks occur when one of the two present domain-wall types reach the N\'eel state, and there are two of them on both sides because they are shifted in opposite directions by the effective chiral fields. }
\end{figure}

In \figref{fig:dmi_supp:DWR_contributions}, we have plotted the various contributions that make up the fitted curves in Figure 3(c) (main text). The purple solid line indicates the intrinsic DW resistance as a function of in-plane field, which gives the same result for $H_x$ and $H_y$ fields. The dark blue dotted line shows the modeled contribution from the AMR effect under the influence of $H_x$ fields. The light blue dash-dotted line shows a calculation of what the AMR effect would look like if we would not take into account the magnetization canting: the AMR resistance simply saturates at high $H_x$. The contribution from AMR as a function of $H_y$ (dashed green curve) is quite small, and reduces at higher in-plane fields since the DW loses its slight N\'eel character. Note that, since we always measure DWs of both polarities in experiment, the modeled AMR under $H_x$ fields is a superposition of two curves, mutually shifted by the chiral field $\frac{2}{\pi} H_{D}$.

\section*{Supplementary References}

\clearpage
\section*{Supplementary Figures}

\begin{thebibliography}{33}
\expandafter\ifx\csname url\endcsname\relax
  \def\url#1{\texttt{#1}}\fi
\expandafter\ifx\csname urlprefix\endcsname\relax\def\urlprefix{URL }\fi
\providecommand{\bibinfo}[2]{#2}
\providecommand{\eprint}[2][]{\url{#2}}

\bibitem{Thiaville2012}
\bibinfo{author}{Thiaville, A.}, \bibinfo{author}{Rohart, S.},
  \bibinfo{author}{Ju\'{e}, E.}, \bibinfo{author}{Cros, V.} \&
  \bibinfo{author}{Fert, A.}
\newblock \bibinfo{title}{{Dynamics of Dzyaloshinskii domain walls in ultrathin
  magnetic films}}.
\newblock \emph{\bibinfo{journal}{Europhys. Lett.}}
  \textbf{\bibinfo{volume}{100}}, \bibinfo{pages}{57002}
  (\bibinfo{year}{2012}).

\bibitem{Emori2013a}
\bibinfo{author}{Emori, S.}, \bibinfo{author}{Bauer, U.}, \bibinfo{author}{Ahn,
  S.-M.}, \bibinfo{author}{Martinez, E.} \& \bibinfo{author}{Beach, G. S.~D.}
\newblock \bibinfo{title}{{Current-driven dynamics of chiral ferromagnetic
  domain walls}}.
\newblock \emph{\bibinfo{journal}{Nat. Mater.}} \textbf{\bibinfo{volume}{12}},
  \bibinfo{pages}{611--6} (\bibinfo{year}{2013}).

\bibitem{Ryu2013}
\bibinfo{author}{Ryu, K.-S.}, \bibinfo{author}{Thomas, L.},
  \bibinfo{author}{Yang, S.-H.} \& \bibinfo{author}{Parkin, S.}
\newblock \bibinfo{title}{{Chiral spin torque at magnetic domain walls}}.
\newblock \emph{\bibinfo{journal}{Nat. Nanotechnol.}}
  \textbf{\bibinfo{volume}{8}}, \bibinfo{pages}{527--33}
  (\bibinfo{year}{2013}).

\bibitem{Je2013}
\bibinfo{author}{Je, S.-G.} \emph{et~al.}
\newblock \bibinfo{title}{{Asymmetric magnetic domain-wall motion by the
  Dzyaloshinskii-Moriya interaction}}.
\newblock \emph{\bibinfo{journal}{Phys. Rev. B}} \textbf{\bibinfo{volume}{88}},
  \bibinfo{pages}{214401} (\bibinfo{year}{2013}).

\bibitem{Parkin2008}
\bibinfo{author}{Parkin, S. S.~P.}, \bibinfo{author}{Hayashi, M.} \&
  \bibinfo{author}{Thomas, L.}
\newblock \bibinfo{title}{{Magnetic domain-wall racetrack memory}}.
\newblock \emph{\bibinfo{journal}{Science}} \textbf{\bibinfo{volume}{320}},
  \bibinfo{pages}{190--4} (\bibinfo{year}{2008}).

\bibitem{Thiaville2005}
\bibinfo{author}{Thiaville, A.}, \bibinfo{author}{Nakatani, Y.},
  \bibinfo{author}{Miltat, J.} \& \bibinfo{author}{Suzuki, Y.}
\newblock \bibinfo{title}{{Micromagnetic understanding of current-driven domain
  wall motion in patterned nanowires}}.
\newblock \emph{\bibinfo{journal}{Europhys. Lett.}}
  \textbf{\bibinfo{volume}{69}}, \bibinfo{pages}{990} (\bibinfo{year}{2005}).

\bibitem{Miron2011a}
\bibinfo{author}{Miron, I.~M.} \emph{et~al.}
\newblock \bibinfo{title}{{Perpendicular switching of a single ferromagnetic
  layer induced by in-plane current injection}}.
\newblock \emph{\bibinfo{journal}{Nature}} \textbf{\bibinfo{volume}{476}},
  \bibinfo{pages}{189--194} (\bibinfo{year}{2011}).

\bibitem{Liu2012}
\bibinfo{author}{Liu, L.} \emph{et~al.}
\newblock \bibinfo{title}{{Spin-Torque Switching with the Giant Spin Hall
  Effect of Tantalum}}.
\newblock \emph{\bibinfo{journal}{Science}} \textbf{\bibinfo{volume}{336}},
  \bibinfo{pages}{555--558} (\bibinfo{year}{2012}).

\bibitem{Garello2013}
\bibinfo{author}{Garello, K.} \emph{et~al.}
\newblock \bibinfo{title}{{Symmetry and magnitude of spin-orbit torques in
  ferromagnetic heterostructures}}.
\newblock \emph{\bibinfo{journal}{Nat. Nanotechnol.}}
  \textbf{\bibinfo{volume}{8}}, \bibinfo{pages}{587--93}
  (\bibinfo{year}{2013}).

\bibitem{Miron2011}
\bibinfo{author}{Miron, I.~M.} \emph{et~al.}
\newblock \bibinfo{title}{{Fast current-induced domain-wall motion controlled
  by the Rashba effect}}.
\newblock \emph{\bibinfo{journal}{Nat. Mater.}} \textbf{\bibinfo{volume}{10}},
  \bibinfo{pages}{419--23} (\bibinfo{year}{2011}).

\bibitem{Wang2012a}
\bibinfo{author}{Wang, X.} \& \bibinfo{author}{Manchon, A.}
\newblock \bibinfo{title}{{Diffusive Spin Dynamics in Ferromagnetic Thin Films
  with a Rashba Interaction}}.
\newblock \emph{\bibinfo{journal}{Phys. Rev. Lett.}}
  \textbf{\bibinfo{volume}{108}}, \bibinfo{pages}{117201}
  (\bibinfo{year}{2012}).

\bibitem{Kim2012b}
\bibinfo{author}{Kim, K.-W.}, \bibinfo{author}{Seo, S.-M.},
  \bibinfo{author}{Ryu, J.}, \bibinfo{author}{Lee, K.-J.} \&
  \bibinfo{author}{Lee, H.-W.}
\newblock \bibinfo{title}{{Magnetization dynamics induced by in-plane currents
  in ultrathin magnetic nanostructures with Rashba spin-orbit coupling}}.
\newblock \emph{\bibinfo{journal}{Phys. Rev. B}} \textbf{\bibinfo{volume}{85}},
  \bibinfo{pages}{180404(R)} (\bibinfo{year}{2012}).

\bibitem{Hirsch1999}
\bibinfo{author}{Hirsch, J.}
\newblock \bibinfo{title}{{Spin Hall Effect}}.
\newblock \emph{\bibinfo{journal}{Phys. Rev. Lett.}}
  \textbf{\bibinfo{volume}{83}}, \bibinfo{pages}{1834--1837}
  (\bibinfo{year}{1999}).

\bibitem{Haazen2013}
\bibinfo{author}{Haazen, P. P.~J.} \emph{et~al.}
\newblock \bibinfo{title}{{Domain wall depinning governed by the spin Hall
  effect}}.
\newblock \emph{\bibinfo{journal}{Nat. Mater.}} \textbf{\bibinfo{volume}{12}},
  \bibinfo{pages}{299} (\bibinfo{year}{2013}).

\bibitem{Khvalkovskiy2013}
\bibinfo{author}{Khvalkovskiy, A.~V.} \emph{et~al.}
\newblock \bibinfo{title}{{Matching domain-wall configuration and spin-orbit
  torques for efficient domain-wall motion}}.
\newblock \emph{\bibinfo{journal}{Phys. Rev. B}} \textbf{\bibinfo{volume}{87}},
  \bibinfo{pages}{020402} (\bibinfo{year}{2013}).

\bibitem{Moore2008}
\bibinfo{author}{Moore, T.~A.} \emph{et~al.}
\newblock \bibinfo{title}{{High domain wall velocities induced by current in
  ultrathin Pt/Co/AlOx wires with perpendicular magnetic anisotropy}}.
\newblock \emph{\bibinfo{journal}{Appl. Phys. Lett.}}
  \textbf{\bibinfo{volume}{93}}, \bibinfo{pages}{262504}
  (\bibinfo{year}{2008}).

\bibitem{Koyama2013}
\bibinfo{author}{Koyama, T.} \emph{et~al.}
\newblock \bibinfo{title}{{Current-Induced Magnetic Domain Wall Motion in a
  Co/Ni Nanowire with Structural Inversion Asymmetry}}.
\newblock \emph{\bibinfo{journal}{Appl. Phys. Express}}
  \textbf{\bibinfo{volume}{6}}, \bibinfo{pages}{033001} (\bibinfo{year}{2013}).

\bibitem{Emori2013b}
\bibinfo{author}{Emori, S.}, \bibinfo{author}{Martinez, E.} \&
  \bibinfo{author}{Bauer, U.}
\newblock \bibinfo{title}{{Spin Hall torque magnetometry of Dzyaloshinskii
  domain walls}}.
\newblock \emph{\bibinfo{journal}{arXiv}} \textbf{\bibinfo{volume}{1308.1432}}
  (\bibinfo{year}{2013}).

\bibitem{Torrejon2013}
\bibinfo{author}{Torrejon, J.} \emph{et~al.}
\newblock \bibinfo{title}{{Interface control of the magnetic chirality in
  TaN|CoFeB|MgO heterosctructures}}.
\newblock \emph{\bibinfo{journal}{arXiv}} \textbf{\bibinfo{volume}{1308.1751}}
  (\bibinfo{year}{2013}).

\bibitem{Moriya1960}
\bibinfo{author}{Moriya, T.}
\newblock \bibinfo{title}{{New mechanism of anisotropic superexchange
  interaction}}.
\newblock \emph{\bibinfo{journal}{Phys. Rev. Lett.}}
  \textbf{\bibinfo{volume}{4}}, \bibinfo{pages}{228} (\bibinfo{year}{1960}).

\bibitem{Heide2008}
\bibinfo{author}{Heide, M.}, \bibinfo{author}{Bihlmayer, G.} \&
  \bibinfo{author}{Bl\"{u}gel, S.}
\newblock \bibinfo{title}{{Dzyaloshinskii-Moriya interaction accounting for the
  orientation of magnetic domains in ultrathin films: Fe/W(110)}}.
\newblock \emph{\bibinfo{journal}{Phys. Rev. B}} \textbf{\bibinfo{volume}{78}},
  \bibinfo{pages}{140403} (\bibinfo{year}{2008}).

\bibitem{Chen2013}
\bibinfo{author}{Chen, G.} \emph{et~al.}
\newblock \bibinfo{title}{{Novel Chiral Magnetic Domain Wall Structure in
  Fe/Ni/Cu(001) Films}}.
\newblock \emph{\bibinfo{journal}{Phys. Rev. Lett.}}
  \textbf{\bibinfo{volume}{110}}, \bibinfo{pages}{177204}
  (\bibinfo{year}{2013}).

\bibitem{Chen2013a}
\bibinfo{author}{Chen, G.} \emph{et~al.}
\newblock \bibinfo{title}{{Tailoring the chirality of magnetic domain walls by
  interface engineering}}.
\newblock \emph{\bibinfo{journal}{Nat. Commun.}} \textbf{\bibinfo{volume}{4}},
  \bibinfo{pages}{2671} (\bibinfo{year}{2013}).

\bibitem{Franken2011}
\bibinfo{author}{Franken, J.~H.} \emph{et~al.}
\newblock \bibinfo{title}{{Precise control of domain wall injection and pinning
  using helium and gallium focused ion beams}}.
\newblock \emph{\bibinfo{journal}{J. Appl. Phys.}}
  \textbf{\bibinfo{volume}{109}}, \bibinfo{pages}{07D504}
  (\bibinfo{year}{2011}).

\bibitem{Franken2012b}
\bibinfo{author}{Franken, J.~H.}, \bibinfo{author}{Hoeijmakers, M.},
  \bibinfo{author}{Swagten, H. J.~M.} \& \bibinfo{author}{Koopmans, B.}
\newblock \bibinfo{title}{{Tunable Resistivity of Individual Magnetic Domain
  Walls}}.
\newblock \emph{\bibinfo{journal}{Phys. Rev. Lett.}}
  \textbf{\bibinfo{volume}{108}}, \bibinfo{pages}{037205}
  (\bibinfo{year}{2012}).

\bibitem{Aziz2006}
\bibinfo{author}{Aziz, A.} \emph{et~al.}
\newblock \bibinfo{title}{{Angular Dependence of Domain Wall Resistivity in
  Artificial Magnetic Domain Structures}}.
\newblock \emph{\bibinfo{journal}{Phys. Rev. Lett.}}
  \textbf{\bibinfo{volume}{97}}, \bibinfo{pages}{206602}
  (\bibinfo{year}{2006}).

\bibitem{Levy1997}
\bibinfo{author}{Levy, P.~M.} \& \bibinfo{author}{Zhang, S.}
\newblock \bibinfo{title}{{Resistivity due to Domain Wall Scattering}}.
\newblock \emph{\bibinfo{journal}{Phys. Rev. Lett.}}
  \textbf{\bibinfo{volume}{79}}, \bibinfo{pages}{5110--5113}
  (\bibinfo{year}{1997}).

\bibitem{Koyama2011}
\bibinfo{author}{Koyama, T.} \emph{et~al.}
\newblock \bibinfo{title}{{Observation of the intrinsic pinning of a magnetic
  domain wall in a ferromagnetic nanowire}}.
\newblock \emph{\bibinfo{journal}{Nat. Mater.}} \textbf{\bibinfo{volume}{10}},
  \bibinfo{pages}{194--197} (\bibinfo{year}{2011}).

\bibitem{Franken2012}
\bibinfo{author}{Franken, J.~H.}, \bibinfo{author}{Hoeijmakers, M.},
  \bibinfo{author}{Lavrijsen, R.} \& \bibinfo{author}{Swagten, H. J.~M.}
\newblock \bibinfo{title}{{Domain-wall pinning by local control of anisotropy
  in Pt/Co/Pt strips}}.
\newblock \emph{\bibinfo{journal}{J. Phys. Cond. Matter}}
  \textbf{\bibinfo{volume}{24}}, \bibinfo{pages}{024216}
  (\bibinfo{year}{2012}).

\bibitem{Boulle2013}
\bibinfo{author}{Boulle, O.} \emph{et~al.}
\newblock \bibinfo{title}{{Domain Wall Tilting in the Presence of the
  Dzyaloshinskii-Moriya Interaction in Out-of-Plane Magnetized Magnetic
  Nanotracks}}.
\newblock \emph{\bibinfo{journal}{Phys. Rev. Lett.}}
  \textbf{\bibinfo{volume}{111}}, \bibinfo{pages}{217203}
  (\bibinfo{year}{2013}).

\bibitem{Freimuth2013}
\bibinfo{author}{Freimuth, F.}, \bibinfo{author}{Bl\"{u}gel, S.} \&
  \bibinfo{author}{Mokrousov, Y.}
\newblock \bibinfo{title}{{Berry phase theory of Dzyaloshinskii-Moriya
  interaction and spin-orbit torques}}.
\newblock \emph{\bibinfo{journal}{arXiv}} \textbf{\bibinfo{volume}{1308.5983}}
  (\bibinfo{year}{2013}).

\bibitem{Bandiera2011}
\bibinfo{author}{Bandiera, S.}, \bibinfo{author}{Sousa, R.~R.},
  \bibinfo{author}{Rodmacq, B.~B.} \& \bibinfo{author}{Dieny, B.}
\newblock \bibinfo{title}{{Asymmetric Interfacial Perpendicular Magnetic
  Anisotropy in Pt/Co/Pt Trilayers}}.
\newblock \emph{\bibinfo{journal}{IEEE Magn. Lett.}}
  \textbf{\bibinfo{volume}{2}}, \bibinfo{pages}{3000504}
  (\bibinfo{year}{2011}).

\bibitem{Nguyen2011a}
\bibinfo{author}{Nguyen, V.~D.} \emph{et~al.}
\newblock \bibinfo{title}{{Detection of Domain-Wall Position and Magnetization
  Reversal in Nanostructures Using the Magnon Contribution to the
  Resistivity}}.
\newblock \emph{\bibinfo{journal}{Phys. Rev. Lett.}}
  \textbf{\bibinfo{volume}{107}}, \bibinfo{pages}{136605}
  (\bibinfo{year}{2011}).

\end{thebibliography}

\begin{thebibliography}{1}
\expandafter\ifx\csname url\endcsname\relax
  \def\url#1{\texttt{#1}}\fi
\expandafter\ifx\csname urlprefix\endcsname\relax\def\urlprefix{URL }\fi
\providecommand{\bibinfo}[2]{#2}
\providecommand{\eprint}[2][]{\url{#2}}

\bibitem{Haazen2013}
\bibinfo{author}{Haazen, P. P.~J.} \emph{et~al.}
\newblock \bibinfo{title}{{Domain wall depinning governed by the spin Hall
  effect}}.
\newblock \emph{\bibinfo{journal}{Nat. Mater.}} \textbf{\bibinfo{volume}{12}},
  \bibinfo{pages}{299} (\bibinfo{year}{2013}).

\bibitem{Ryu2013}
\bibinfo{author}{Ryu, K.-S.}, \bibinfo{author}{Thomas, L.},
  \bibinfo{author}{Yang, S.-H.} \& \bibinfo{author}{Parkin, S.}
\newblock \bibinfo{title}{{Chiral spin torque at magnetic domain walls}}.
\newblock \emph{\bibinfo{journal}{Nat. Nanotechnol.}}
  \textbf{\bibinfo{volume}{8}}, \bibinfo{pages}{527--33}
  (\bibinfo{year}{2013}).

\bibitem{Thiaville2012}
\bibinfo{author}{Thiaville, A.}, \bibinfo{author}{Rohart, S.},
  \bibinfo{author}{Ju\'{e}, E.}, \bibinfo{author}{Cros, V.} \&
  \bibinfo{author}{Fert, A.}
\newblock \bibinfo{title}{{Dynamics of Dzyaloshinskii domain walls in ultrathin
  magnetic films}}.
\newblock \emph{\bibinfo{journal}{Europhys. Lett.}}
  \textbf{\bibinfo{volume}{100}}, \bibinfo{pages}{57002}
  (\bibinfo{year}{2012}).

\bibitem{Liu2011a}
\bibinfo{author}{Liu, L.}, \bibinfo{author}{Moriyama, T.},
  \bibinfo{author}{Ralph, D.~C.} \& \bibinfo{author}{Buhrman, R.~A.}
\newblock \bibinfo{title}{{Spin-Torque Ferromagnetic Resonance Induced by the
  Spin Hall Effect}}.
\newblock \emph{\bibinfo{journal}{Phys. Rev. Lett.}}
  \textbf{\bibinfo{volume}{106}}, \bibinfo{pages}{036601}
  (\bibinfo{year}{2011}).

\bibitem{Liu}
\bibinfo{author}{Liu, L.}, \bibinfo{author}{Buhrman, R.~A.} \&
  \bibinfo{author}{Ralph, D.~C.}
\newblock \bibinfo{title}{{Review and Analysis of Measurements of the Spin Hall
  Effect in Platinum}}.
\newblock \emph{\bibinfo{journal}{arXiv}} \textbf{\bibinfo{volume}{1111.3702}}
  (\bibinfo{year}{2011}).

\bibitem{Levy1997}
\bibinfo{author}{Levy, P.~M.} \& \bibinfo{author}{Zhang, S.}
\newblock \bibinfo{title}{{Resistivity due to Domain Wall Scattering}}.
\newblock \emph{\bibinfo{journal}{Phys. Rev. Lett.}}
  \textbf{\bibinfo{volume}{79}}, \bibinfo{pages}{5110--5113}
  (\bibinfo{year}{1997}).

\bibitem{Franken2012b}
\bibinfo{author}{Franken, J.~H.}, \bibinfo{author}{Hoeijmakers, M.},
  \bibinfo{author}{Swagten, H. J.~M.} \& \bibinfo{author}{Koopmans, B.}
\newblock \bibinfo{title}{{Tunable Resistivity of Individual Magnetic Domain
  Walls}}.
\newblock \emph{\bibinfo{journal}{Phys. Rev. Lett.}}
  \textbf{\bibinfo{volume}{108}}, \bibinfo{pages}{037205}
  (\bibinfo{year}{2012}).

\bibitem{Franken2012}
\bibinfo{author}{Franken, J.~H.}, \bibinfo{author}{Hoeijmakers, M.},
  \bibinfo{author}{Lavrijsen, R.} \& \bibinfo{author}{Swagten, H. J.~M.}
\newblock \bibinfo{title}{{Domain-wall pinning by local control of anisotropy
  in Pt/Co/Pt strips}}.
\newblock \emph{\bibinfo{journal}{J. Phys. Cond. Matter}}
  \textbf{\bibinfo{volume}{24}}, \bibinfo{pages}{024216}
  (\bibinfo{year}{2012}).

\bibitem{Kobs2011}
\bibinfo{author}{Kobs, A.} \emph{et~al.}
\newblock \bibinfo{title}{{Anisotropic Interface Magnetoresistance in Pt/Co/Pt
  Sandwiches}}.
\newblock \emph{\bibinfo{journal}{Phys. Rev. Lett.}}
  \textbf{\bibinfo{volume}{106}}, \bibinfo{pages}{217207}
  (\bibinfo{year}{2011}).

\end{thebibliography}
\end{document}